\documentstyle[12pt]{article}
\textheight
210mm
\evensidemargin
-0.5cm
\begin{document}
\begin{flushright}
IITAP-95-01 \\
INR-0891/95 \\
May  1995
\end{flushright}
\begin{center}
{\large \bf BFKL QCD Pomeron in
High Energy Hadron Collisions: \\
 Inclusive Dijet Production}
\\
\vspace{0.7cm}
{\large  Victor T. Kim${}^{\dagger}$
\footnote{\em
e-mail: $kim@fnpnpi.pnpi.spb.ru$}
and Grigorii B.
Pivovarov${}^{\ddagger}$
\footnote{\em e-mail:
$gbpivo@ms2.inr.ac.ru$} }\\
\vspace*{0.5cm}
{\em ${}^\dagger$ :
International Institute of Theoretical and
Applied Physics, \\
Iowa
State University, Ames, Iowa 50011-3160 \\
and \\
St.Petersburg
Nuclear Physics Institute,
 188350 Gatchina,
Russia}\\
\vspace*{0.3cm}
{\em ${}^\ddagger$ :
Institute for Nuclear
Research, 117312 Moscow,
Russia}
\end{center}

\vspace*{0.5cm}
\begin{center}
{\large \bf
Abstract}
\end{center}

We calculate inclusive dijet production cross section in
high energy hadron collisions
within the BFKL resummation formalism
for the
QCD Pomeron. Unlike the previous calculations with the Pomeron
developing
only between tagging jets, we include also the Pomerons
which are
adjacent to the hadrons. With these adjacent
Pomerons we  define a new
object --- the BFKL structure function of hadron --- which
enables one to calculate
the inclusive dijet production for any rapidity intervals.
We present predictions for the K-factor and
the azimuthal angle decorrelation
in the inclusive dijet production for
Fermilab-Tevatron and CERN-LHC energies.

\vspace*{0.5cm}
PACS
number(s): 13.87.Ce,12.38.Cy,13.85.Hd

\vspace*{0.5cm}

\newpage

At present, much attention is being paid to the perturbative
QCD
Pomeron obtained by Balitsky, Fadin, Kuraev and Lipatov (BFKL)
\cite{Lip76}.
One of the reasons is that it relates hard processes
($
-t = Q^2 \gg {\Lambda^2_{QCD}}$) and
semi-hard ones ($s \gg -t =
Q^2 \gg {\Lambda^2_{QCD}}$):
It sums up leading energy logarithms of
perturbative QCD
into a singularity in the complex angular momentum
plane.
Several proposals to find direct manifestations of the BFKL
Pomeron
are available in the literature, see, e.g.,
[2-9], but
it is still difficult to get the necessary experimental data.

Among those
proposals the first one was made by Mueller and
Navelet \cite{Mue87}.
They pointed out that the inclusive dijet
production in high energy
hadron collisions may serve as a
probe for the BFKL Pomeron. Namely,
a specific exponential growth
of the cross section K-factor with the
rapidity interval of
tagged jets was predicted.

This idea was
further developed in \cite{Del94,Sti94},
where it was found out that
the relevant object
may be the azimuthal angle correlation of
jets.

Note, that these studies were restricted to a consideration
of
some special configuration of the inclusive dijet production.
Namely,
in \cite{Mue87,Del94,Sti94} only the production
cross section of the most
forward and most backward jets
is considered (Fig. 1(a)). However,
unlike the usual hard QCD processes
with strong $k_\perp$-ordering,
in BFKL Pomeron kinematics \cite{Gri83} there
are strong rapidity ordering and weak $k_\perp$-diffusion.
This means, in particular, that the most forward/backward
jets do not need to have the hardest $k_\perp$.
Therefore, there is no warranty that one can tag
most
forward/backward jets without a dedicated full-acceptance
detector \cite{Bjo92}.

Unfortunately, the available detectors (CDF and
D$\emptyset$)
at the  Fermilab Tevatron as well as forthcoming detectors
at the CERN
Large Hadron Collider (LHC) have limited acceptance in
$k_\perp$
(${k_\perp}_{min} \sim$ tens GeV/c) and (pseudo) rapidity
for tagging jets. So, one cannot compare results of
\cite{Mue87,Del94,Sti94}
with the preliminary D$\emptyset$ data
on
the large rapidity interval dijets \cite{Heu94}
without an analysis
of jet radiation
beyond the acceptance of the detectors.

In this
paper we study, within the BFKL approach, the
inclusive dijet cross
section in high energy hadron collisions
without any restrictions on
untagging jets. Our results provide an opportunity
to confront BFKL Pomeron predictions on the inclusive dijet
production with data
which could be extracted
from the existing CDF and D$\emptyset$ jet event
samples
after a modification of the jet analysis
algorithms. This may be decisive in checking applicability
of the factorization hypothesis for high energy hadron collisions
(arguments against it see in \cite{Bra93}) and
the leading logarithm approximation.

Removing the restriction on tagging jets to be most
forward/backward
one should take into account
additional
contributions to the cross section with
jets more rapid than the
tagging ones.
There are three such contributions:
two with a couple
of Pomerons
(Figs. 1(b),1(c)) and one with three (Fig. 1(d)).
In
this paper, we will call the Pomerons developing
between
colliding hadrons and their descendant jets the
adjacent
Pomerons and the Pomeron developing
between the tagging jets
the
inner Pomeron. These additional contributions
contain extra power
of $\alpha_S$ per extra Pomeron but hardly
could they be regarded as
corrections since they
are also proportional to a
kinematically
dependent factor which one can loosely treat as
the number
of partons in the hadron moving faster
than the descendant tagging
jet.

Contribution to the cross section of Fig. 1(a)
considered by
Mueller and Navelet \cite{Mue87}
is
\begin{equation}
\frac{x_1x_2d\sigma_{\{ P
\}}}{dx_1dx_2d^{2}k_{1\perp}d^{2}k_{2\perp}} =
\frac{\alpha_{S}
C_A}{k^2_{1\perp}}\frac{\alpha_{S}
C_A}{k^2_{2\perp}}
x_1F_A(x_1,\mu^2_1)x_2F_B(x_2,\mu^2_2)
f^{BFKL}(k_{1\perp},k_{2\perp},y),
\label{mueller}
\end{equation}
where the subscript on $\sigma_{\{P\}}$ labels the contribution
to the cross section as a single-Pomeron; $C_A=3$ is a color
group factor; $x_i$ are the longitudinal momentum fractions
of the tagging jets;
$k_{i\perp}$ are the transverse
momenta; $F_{A,B}$ are the effective structure
functions of colliding hadrons;
$y = \ln(x_1x_2s/k_{1\perp}k_{2\perp})$ is the relative rapidity
of tagging jets
and, finally, $f^{BFKL}$
is the solution for the BFKL equation.

If one retains in $f^{BFKL}(k_{1\perp},k_{2\perp},y)$
of Eq.~(\ref{mueller}) only the leading $\alpha_S$-independent
contribution to its $\alpha_S$-expansion, one gets
the result of \cite{Com84}:
\begin{equation}
\frac{x_1x_2d\sigma_{\{P\}}}{dx_1dx_2d^{2}k_{1\perp}d^{2}k_{2\perp}} =
\frac{\alpha_{S} C_A}{k^2_{1\perp}}
\frac{\alpha_{S} C_A}{k^2_{2\perp}}
x_1F_A(x_1,\mu^2_1)x_2F_B(x_2,\mu^2_2)
\frac{1}{2}\delta^{(2)}(k_{1\perp}-k_{2\perp}),
\label{com}
\end{equation}
(An interesting feature of
this result is that it does not contain specific
contributions from gluons and quarks --- they turn out
to be the same up to a simple group factor in the
high energy limit. This provides the possibility
to hide all the nonperturbative physics in
the pair of effective structure functions $F_{A,B}=
G_{A,B}+\frac{C_F}{C_A}Q_{A,B}+\frac{C_F}{C_A}\overline Q_{A,B}$.) What
distinguishes the cross section of Eq.~(\ref{mueller})
from the analogous one of Eq.~(\ref{com}) is a systematic
resummation of $(\alpha_S y)$-corrections to the hard subprocess
cross section, which is necessary when the relative rapidity
of jets, $y$, is not small.

The solution for the BFKL equation has the following
integral representation \cite{Lip76}:
\begin{equation}
f^{BFKL}(k_{1 \perp},k_{2 \perp},y)=
\sum_{n=-\infty}^{\infty}\int d\nu
\chi_{n,\nu}(k_{1 \perp})e^{y\omega(n,\nu)}\chi_{n,\nu}^*(k_{2 \perp}),
\label{lipatov}
\end{equation}
where the star means complex conjugation;
\begin{equation}
\chi_{n,\nu}(k_{\perp})=\frac{(k_{\perp}^2)^{-\frac{1}{2}+i\nu}
e^{in\phi}}{2\pi}
\label{eigenfunction}
\end{equation}
are Lipatov's eigenfunctions and
$$\omega(n,\nu) = \frac{2 \alpha_{S} C_A}{\pi}
\biggl[ \psi(1) - Re \, \psi \biggl( \frac{|n|+1}{2} +
i\nu \biggr) \biggr]$$ are
Lipatov's eigenvalues. Here $\psi$ is the logarithmic derivative
of Euler Gamma-function.

Making use of the above-introduced objects
we rewrite  Eq.(\ref{mueller}) as
 follows:
\begin{equation}
\frac{x_1x_2d\sigma_{\{P\}}}{dx_1dx_2d^{2}k_{1 \perp}d^{2}k_{2 \perp}}  =
\frac{\alpha_{S} C_A}{k^2_{1 \perp}}\frac{\alpha_{S} C_A}{k^2_{2 \perp}}
x_1x_2
\sum_n\int d\nu
F_A(x_1,\mu^2_1)
\left[
\chi_{n,\nu}(k_{1 \perp})
e^{y\omega(n,\nu)}
\chi_{n,\nu}^*(k_{2 \perp})
\right]
F_B(x_2,\mu^2_2).
\label{cast}
\end{equation}

   As one can guess, subprocesses of Fig. 1(b)-1(d) with the
adjacent Pomerons contribute to the
 effective structure
functions, i.e., one can account for them by just adding some
``radiation corrections''  to the structure functions
of Eq.(\ref{cast}):
\begin{eqnarray}
F_A(x_1,\mu^2_1) & \Rightarrow & \Phi_{A}(x_1,\mu^2_1,n,\nu,k_{1\perp})
 \equiv
F_A(x_1,\mu^2_1)+D_{A}(x_1,\mu^2_1,n,\nu,k_{1\perp}), \\
F_B(x_2,\mu^2_2)&\Rightarrow&\Phi_{B}^{\ast}(x_2,\mu^2_2,n,\nu,k_{2\perp})
\equiv
F_B(x_2,\mu^2_2)+D_{B}^{\ast}(x_2,\mu^2_2,n,\nu,k_{2\perp}).
\label{subs}
\end{eqnarray}
The complex conjugation on $\Phi_B$ could be understood
if one look at rhs of Eq.(\ref{cast}) as a matrix element
of a $t$-chanel evolution operator with the relative
rapidity, $y$, as an evolution parameter and $F_B$
as a final state; $(n,\nu)$ are then  ``good quantum numbers''
conserved under the evolution---this makes room for
$(n,\nu)$-dependence of the corrected structure functions.
We note also that the corrected structure functions
may depend on the transverse momenta of the tagging jets.

To get an explicit expression for the radiation correction
to the effective hadron structure functions, let us consider,
for example, the contribution of Fig. 1(b). In terms of the
radiation correction, $D_A$, it is
\begin{eqnarray}
\frac{x_1x_2d\sigma_{\{PP\}A}}{dx_1dx_2d^{2}k_{1 \perp}d^{2}k_{2 \perp}} =
\frac{\alpha_{S} C_A}{k^2_{1 \perp}}\frac{\alpha_{S} C_A}{k^2_{2 \perp}}
x_1x_2 \times \nonumber \\
 \sum_n\int d\nu
D_A(x_1,\mu^2_1,n,\nu,k_{1\perp})
\left[
\chi_{n,\nu}(k_{1 \perp})
e^{y\omega(n,\nu)}
\chi_{n,\nu}^*(k_{2 \perp})
\right]
F_B(x_2,\mu^2_2).
\label{dcomp}
\end{eqnarray}
On the other hand, BFKL summation of the leading energy logarithms
yields
\begin{eqnarray}
\label{bfklcomp}
\frac{x_1x_2d\sigma_{\{PP\}A}}{dx_1dx_2d^{2}k_{1\perp}d^{2}k_{2\perp}} =
\frac{\alpha_{S} C_A}{k^2_{1\perp}}\frac{\alpha_{S} C_A}{k^2_{2\perp}}
x_1x_2
\frac{2\alpha_{S} C_A}{\pi^2}\int_{x_1}^1d\xi F_A(\xi,\mu^2_1)
\int_{\mu_{1}}^{\xi\sqrt{s}}\frac{d^{2}q_{1\perp}}{q_{1\perp}^2}
\times\nonumber\\
 \int d^{2}q_{2\perp}f^{BFKL}(q_{1\perp},q_{2\perp},y_1(\xi,q_{1\perp}))
f^{BFKL}(q_{2\perp}+k_{1\perp},k_{2\perp},y)F_B(x_2,\mu^2_2),
\end{eqnarray}
where $\xi, q_{1\perp}$ parameterize the momentum of the most rapid
untagging
 jet moving in the same direction as hadron $A$ (see Fig. 1(a));
\begin{equation}
\label{rapidity}
y_1(\xi,q_{1\perp}) = \log \frac{\xi k_{1\perp}}{x_1 q_{1\perp}}
\end{equation}
is the relative rapidity measuring the rapidity lapse spanned
by the adjacent Pomeron. We  cut the infrared-divergent
transverse momentum integration by the normalization
point, $\mu_1$, of $F_A$. Other integration cuts in Eq.(\ref{bfklcomp}) are
evident from the kinematics\footnote{We consider here only a
kinematically simple case when the rapidities of the tagging jets
have different signs in the center-of-mass frame.}.

Splitting the Pomerons into the product of
Lipatov's eigenfunctions and making the transverse momentum
integrations in Eq.(\ref{bfklcomp}) one gets that, first,
Eqs.(\ref{dcomp}),(\ref{bfklcomp}) are compatible
and, second
%!!!!!!!!!!!!!!!!!!!!!!!!!!!!!!!!!!!!!!!!!!!!!!!!!!!!!!!!!!!!!!!!!!!!!!!!
\footnote{A simplified analog of this formula was found
in a triple-jet inclusive production study \cite{Del93}.},
%!!!!!!!!!!!!!!!!!!!!!!!!!!!!!!!!!!!!!!!!!!!!!!!!!!!!!!!!!!!!!!!!!!!!!!!!
\begin{eqnarray}
\label{explicit}
&&D_A(x_1,\mu^2_1,n,\nu,k_{1\perp}) = i\frac{\alpha_S C_A}{\pi^2}
\frac{\Gamma(|n|/2+1/2 + i\nu)}{\Gamma(|n|/2+1/2-i\nu)}
\times \nonumber \\
&&\int^{\infty}_{-\infty}d\lambda
\frac{\Gamma(|n|/2+1-i(\nu-\lambda))}{\Gamma(|n|/2+1+i(\nu-\lambda))}
\frac{\Gamma(1/2-i\lambda)}{\Gamma(1/2+i\lambda)}
\frac{1}{(i|n|/2+\nu-\lambda+i\epsilon)}\times\nonumber\\
&&\int_x^1\frac{d\xi}{x}\left(\frac{\xi}{x}\right)^{\omega(0,\lambda)}
F_A(\xi,\mu^2_1)
\frac{(k_{1\perp}/\mu_1)^{1+\omega(0,\lambda)-2i\lambda}}
{1+\omega(0,\lambda)-2i\lambda}
\Biggl(1-\left(\frac{\mu_{1}}{\xi\sqrt{s}}\right)^
{1+\omega(0,\lambda)-2i\lambda}\Biggr).
\end{eqnarray}
Some comments on the above formula are in order. The $i\epsilon$
defines the right way to deal with the singularity at
$n=\nu - \lambda=0$. The dependence on the
energy $\sqrt{s}$
is weak---we check it  for the Tevatron and LHC energies
performing the numerical calculations
(see below) with and without suppression of the
last factor in Eq.(\ref{explicit}).

To make a local summary, we have got the following compact
and significant form for the dijet production inclusive cross section:
\begin{eqnarray}
&&\frac{x_1x_2d\sigma_{dijet}}{dx_1dx_2d^{2}k_{1 \perp}d^{2}k_{2 \perp}}=
\frac{\alpha_{S} C_A}{k^2_{1 \perp}}\frac{\alpha_{S} C_A}{k^2_{2 \perp}}
x_1x_2 \times \nonumber \\
&&\sum_n\int d\nu
\Phi_A(x_1,\mu^2_1,n,\nu,k_{1\perp})
\left[
\chi_{n,\nu}(k_{1 \perp})
e^{y\omega(n,\nu)}
\chi_{n,\nu}^*(k_{2 \perp})
\right]
\Phi_B^*(x_2,\mu^2_2,n,\nu,k_{2\perp}),
\label{crossphi}
\end{eqnarray}
where the new structure functions $\Phi_{A,B}$ that depend
on Lipatov's quantum numbers $(n,\nu)$---we call them BFKL
structure functions---can be read off Eqs.(\ref{subs}),(\ref{explicit}).

We expect that Eq.~(\ref{crossphi}) may serve as
an example of factorization  matching the Regge limit of QCD.
 It would be interesting to
make contact between the factorization of
Eq.~(\ref{crossphi}) and the $k_{\perp}$-factorization
of \cite{Col91}.

%!!!!!!!!!!!!!!!!!!!!!!!!!!!!!!!!!!!!!!!!!!!!!!!!!!!!!!!!!!!!!!!!!!!!!
Eq.~(\ref{crossphi}) gets simpler for $x$-symmetric dijet
production with $x_1 = x_2 $ after integration
over transverse momenta squared larger than a cut value,
$k^2_{\perp min}=\mu_1^2=\mu_2^2=Q^2$. Thus,
following \cite{Mue87,Del94,Sti94} we consider
\begin{eqnarray}
\label{norm}
&&\int_{k^2_{\perp min}}^{e^{y^{\ast}}k^2_{\perp min}}
  dk^2_{1\perp}dk^2_{2\perp}
\left(
\frac{x_1x_2d\sigma_{dijet}}{dx_1dx_2d^{2}k_{1 \perp}d^{2}k_{2 \perp}}
\right)
_{x_1=x_2}
\equiv
\frac{(\alpha_S C_A)^2}{s}(e^{y^{\ast}}-1)\times\nonumber\\
&&F_A(e^{y^{\ast}/2}k_{\perp min}/\sqrt{s},k^2_{\perp min})
  F_B(e^{y^{\ast}/2}k_{\perp min}/\sqrt{s},k^2_{\perp min})
  \sum_n\frac{e^{in\phi}}{2\pi}
  C_n(y^{\ast},k_{\perp min}).
\end{eqnarray}
%!!!!!!!!!!!!!!!!!!!!!!!!!!!!!!!!!!!!!!!!!!!!!!!!!!!!!!!!!!!!!!!!!!!!
This depends on the azimuthal angle, $\phi$,
between the tagging jets and an effective relative rapidity,
$y^{\ast}\equiv\ln({x_1x_2s}/{k_{\perp min}^2})$
%===================================================================
\footnote{Note that in the leading logarithm approximation
we have $k_{\perp} \sim k_{\perp min}$, and hence
$y \sim y^{\ast} $}.
%===================================================================
We compute the
Fourier coefficients of the azimuthal angle dependence,
$C_n(y^{\ast},k_{\perp min})$.
The normalization of Eq. (\ref{norm}) makes
$C_0(y^{\ast},k_{\perp min})$
equal to the $K$-factor---the ratio of the
cross section integrated over the azimuthal angle,
to the Born one. The exponential growth
of this quantity with rapidity was predicted in \cite{Mue87}.
Another quantity of interest is the average cosine of
the azimuthal angle between tagging jets,
$<\cos(\phi-\pi)>=C_1(y^{\ast},k_{\perp min})/
C_0(y^{\ast},k_{\perp min})$
\cite{Del94,Sti94} ($\phi=\pi$ for back-to-back jets).
We plot our predictions for the $K$-factor in Fig. 2 and
$<\cos(\phi-\pi)>$ in  Fig. 3. The leading
order CTEQ3L structure functions \cite{Lai94} with
$\Lambda^{(5)}_{QCD}=132$ MeV/c have been used.

We point out the following qualitative features of our numerical results:

(i) the contribution of the adjacent Pomerons to the cross section
is significant---up to 60\% (20\%) at
$k_{\perp min}=20$ GeV/c ($50$ GeV/c)
at the Tevatron energy;

(ii) the contribution of the adjacent Pomerons
slowly dies out as $y^{\ast}$ approaches to its  kinematical bounds;

(iii) growth of the energy as well as the
decrease of the lower cutoff on the transverse
momenta  of the tagging jets
causes the contribution of the adjacent Pomerons to be
even more significant;

(iv) the azimuthal angle decorrelation (deviation of
the average cosine from unit) is less sensitive to
the contribution of the adjacent Pomerons;

As it is apparent from our plots, the resummation effects are
significant not only for large rapidity intervals. Thus,
the region of moderate rapidity intervals
seems also to be promising for BFKL Pomeron manifestation searches.

We should note here that the extraction of data on
high-$k_{\perp}$ jets from the event samples in order to
compare them with the BFKL Pomeron predictions
should be different from the algorithms
directed to a comparison
with perturbative QCD predictions for the hard
processes. These algorithms, motivated by the strong
$k_{\perp}$-ordering of the hard QCD regime, employ
hardest-$k_{\perp}$ jet selection
(see, e.g., \cite{Alg94}). It
is doubtful that one
can reconcile these algorithms with the weak $k_{\perp}$-diffusion
and the strong rapidity ordering of the semi-hard QCD regime,
described by the BFKL resummation. We also note
that our predictions should not be compared with the
preliminary data \cite{Heu94} extracted by the most forward/backward
jet selection criterion. Obviously, one should include
for tagging all the registered pairs of jets
(not only the most forward--backward pair) to compare
with our predictions. In particular, to make a comparison
with Figs. 2,3, one should sum up all the registered
$x$-symmetric dijets ($x_1=x_2$)
with transverse momenta harder than $k_{\perp min}$.

 Based on our study we draw a conclusion
 that the adjacent BFKL Pomerons
can play a decisive role in high energy hadron collisions,
as it may be seen in inclusive dijet production.

 We thank E.A.Kuraev and L.N.Lipatov for stimulating discussions.
 We are grateful to A.J.Somme\-rer,
 J.P.Vary, and B.-L.Young for their kind
 hospitality at the International Institute of
 Theoretical and Applied Physics, Ames, Iowa and support.
 V.T.K. is indebted to  S.Ahn, C.L.Kim, A.Petridis,
 J.Qiu, C.R.Scmidt, and
 S.I.Troyan for helpful conversations.
 G.B.P. wishes to thank F.Paccanoni for fruitful discussions and
 hospitality at Padova University.

%\newpage
\begin{center}
{\large \bf Figure Captions}
\end{center}

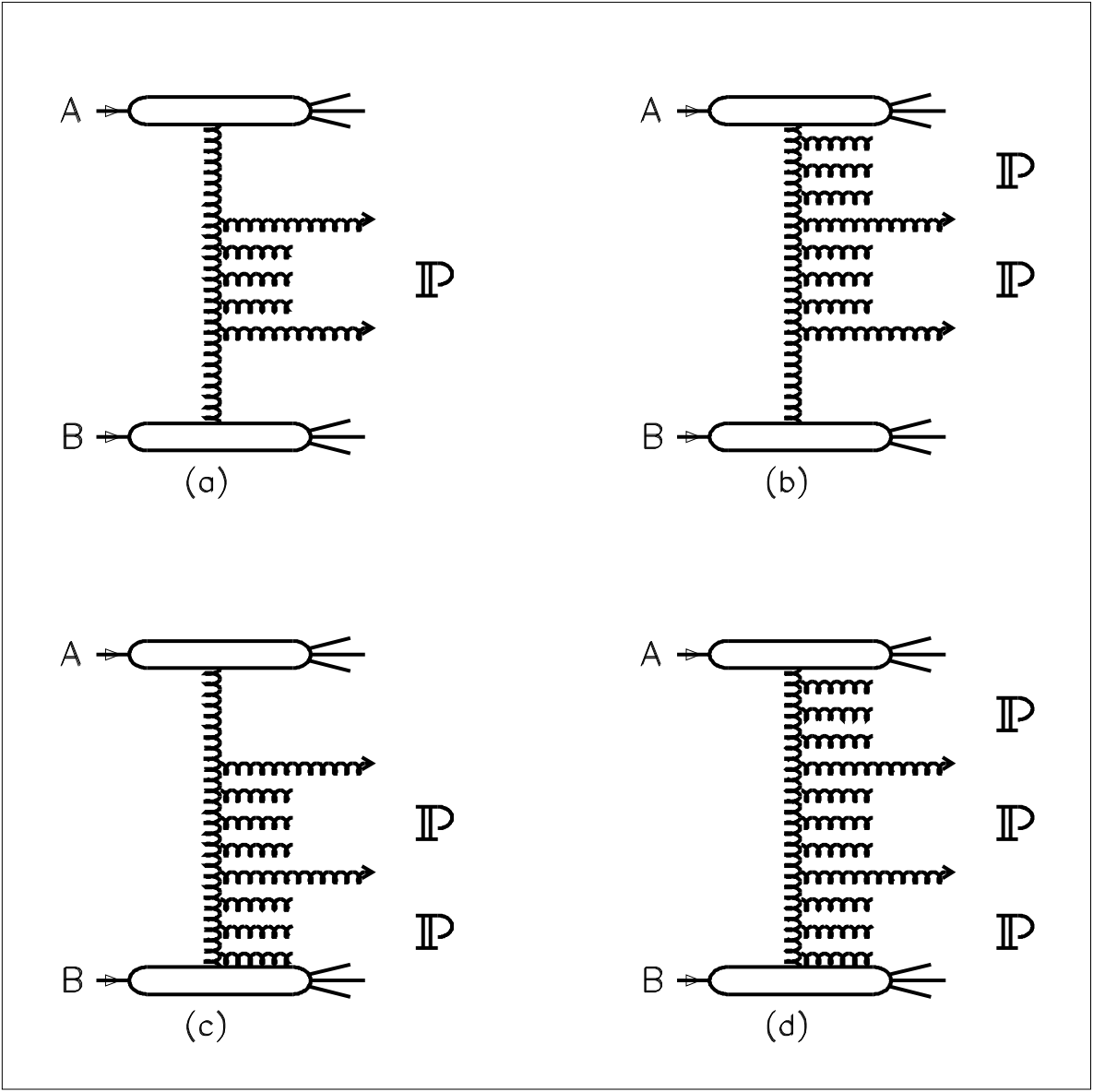
Fig. 1: Subprocesses for the dijet production in a collision
of hadrons $A$ and $B$; vertical curly lines correspond
to the Reggeized gluons; horizontal ones to the real
gluons radiated into the rapidity intervals; arrows
mark gluons producing the tagging jets;
all subprocesses contain the inner Pomeron. (a) subprocess
without the adjacent Pomeron; (b) subprocess with
a Pomeron adjacent to the hadron $A$; (c) same for the
hadron $B$; (d) subprocess with two adjacent Pomerons.

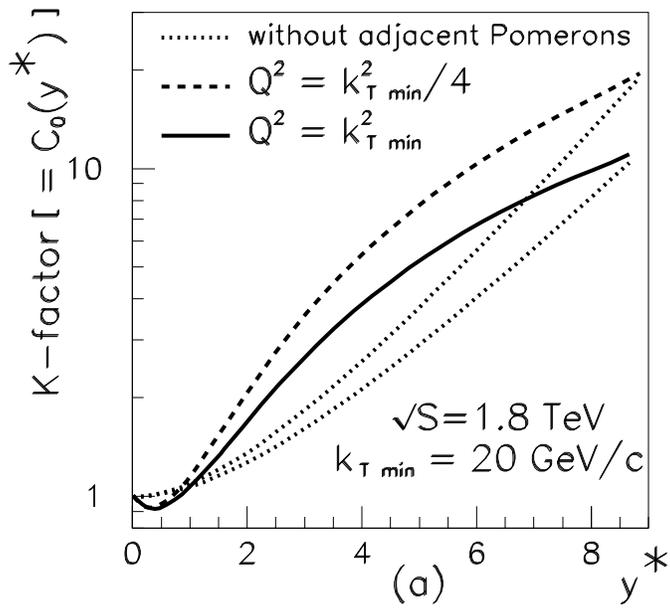
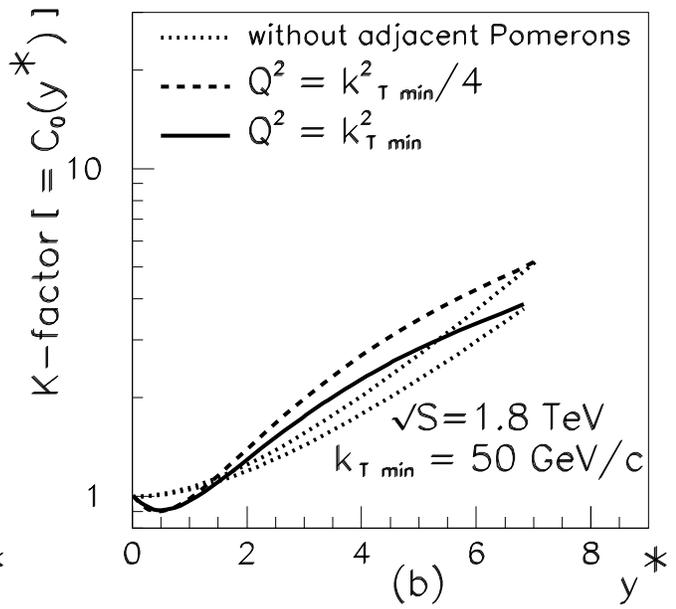
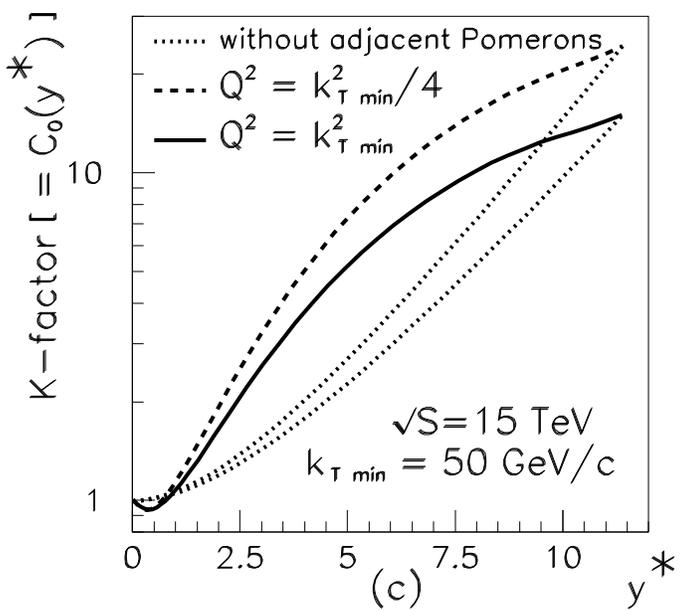
Fig. 2: The $y^{\ast}$-dependence of the dijet K-factor
for various values
of energy, $\sqrt{s}$, transverse momenta cutoff and
normalization point,  $Q^2$, of $\alpha_S$ and $F_{A,B}$.

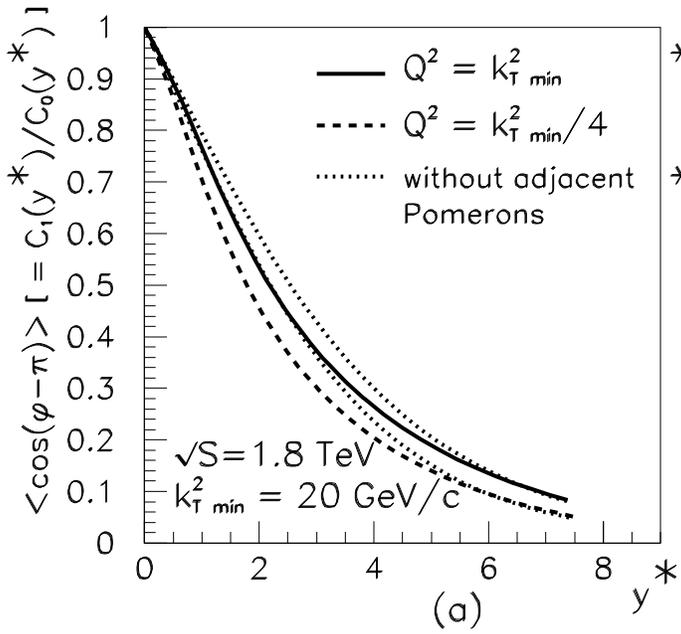
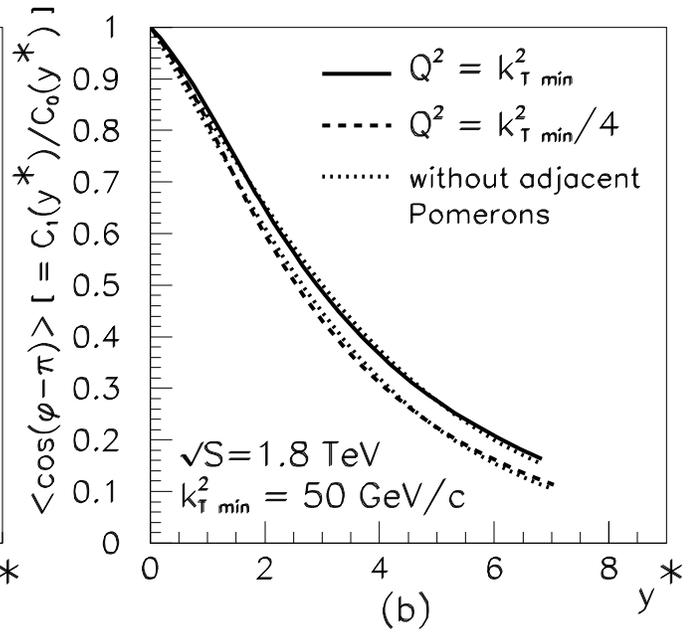
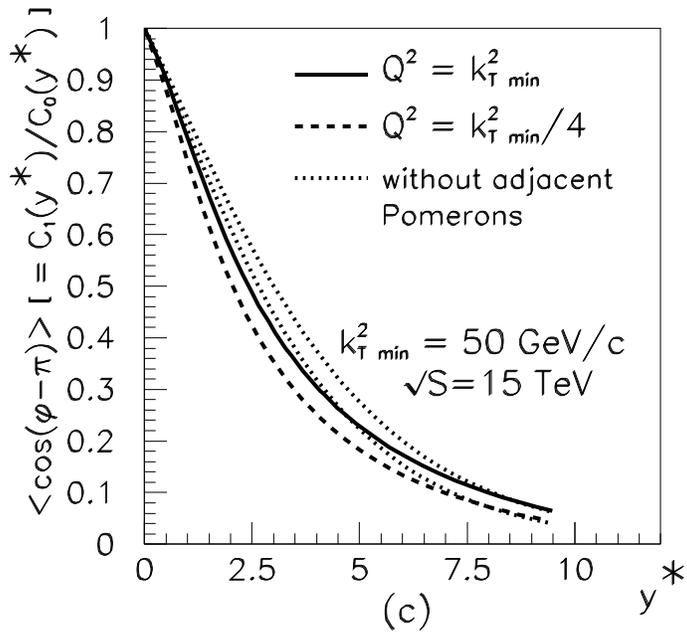
Fig. 3: The $y^{\ast}$-dependence of the average azimuthal angle cosine
between the tagging jets  for various values
of energy, $\sqrt{s}$, transverse momenta cutoff and
normalization point,  $Q^2$, of $\alpha_S$ and $F_{A,B}$.

\end{document}